\documentclass[
 twocolumn,
 secnumarabic
,amssymb, amsmath,nobibnotes, aps, prl]{revtex4}

\begin{document}

\title{On the Definition of Fluctuating Temperature}

\author{B. H. Lavenda}
\email{bernard.lavenda@unicam.it}
\affiliation{Universit\'a degli Studi, Camerino 62032 (MC) Italy}
\date{\today}

\newcommand{\half}{\mbox{\small$\frac{1}{2}$}}
\newcommand{\third}{\mbox{\small$\frac{1}{3}$}}
\newcommand{\twothirds}{\mbox{\small$\frac{2}{3}$}}

\newcommand{\fourth}{\mbox{\small$\frac{1}{4}$}}
\newcommand{\fourthirds}{\mbox{\small$\frac{4}{3}$}}
\newcommand{\fivethirds}{\mbox{\small$\frac{5}{3}$}}
\newcommand{\threehalves}{\mbox{\small$\frac{3}{2}$}}
\newcommand{\summ}{\sum_{i=1}^m\,}
\newcommand{\sumj}{\sum_{j=1}^k\,}
\newcommand{\suml}{\sum_{\ell=1}^m}
\newcommand{\sumll}{\sum_{\ell=k+1}^m}
\newcommand{\ebar}{\bar{E}}
\newcommand{\sumn}{\sum_{i=1}^n\,}
\newcommand{\onem}{\mbox{\small$\frac{1}{m}$}}
\newcommand{\onen}{\mbox{\small$\frac{1}{n}$}}
\newcommand{\mtwo}{\mbox{\small$\frac{m}{2}$}}
\newcommand{\ntwo}{\mbox{\small$\frac{n}{2}$}}
\newcommand{\onenone}{\mbox{\small$\frac{1}{n-1}$}}
\newcommand{\sumntwo}{\sum_{i=2}^n\,}

\begin{abstract}
The Maxwell distribution is derived from the $F$-distribution in the limit where one
of the degrees of freedom of the $\chi^2$ variates tends to infinity. The estimator of
the temperature is consistent, and, hence coincides with the temperature of the heat
reservoir in the asymptotic limit; it is also unbiased and efficient.
Consequently, there is a contradiction between indentifying the Lagrange
multiplier in the variational formalism that Tsallis and co-workers use to maximize his 
nonadditive entropy with respect to escort expectation values in order to derive the
Student $t$- and $r$-distributions and the physical meaning of these variables. 
Only in the asymptotic limit when these distributions become
the $\chi^2$-distributions of MBG  statistics can the Lagrange multiplier be interpreted as the 
inverse temperature. Hence, there is no generalization of the $\chi^2$-distributions that
can be made which involves interpreting the Lagrange multiplier as the inverse 
temperature. The frequency interpretation of the fluctuating temperature is contrasted
with the Bayesian approach that treats a parameter to be estimated as a random variable 
which is equipped with a probability distribution.

\end{abstract}
\maketitle
\flushbottom
Consider a system with $m$ degrees of freedom in thermal contact with a reservoir, at 
temperature $t_0$, with 
$n$ degrees of fredom. The random velocities of the particles in the reservoir, $v_1,\ldots,v_n$,
as well as those of the system, $v_1\ldots,v_m$,   are independent
and normally distributed ($0,t_0)$.\par A \lq fluctuating\rq\ temperature can
be defined as
\begin{equation}
t=\onen\sumn v_i^2 \label{eq:t}
\end{equation}
which becomes the true temperature $t_0$, in the limit as $n\rightarrow\infty$. The 
fluctuating temperature (\ref{eq:t}) is distributed according to $\chi^2$,
\begin{equation}
f_n(t)=\left(\frac{n}{2t_0}\right)^{\ntwo}\frac{t^{\ntwo-1}}{\Gamma(\ntwo)}e^{-nt/2t_0},
\label{eq:t-dis}
\end{equation}
whose moments are given in terms of the temperature of the heat bath, $t_0$.\par
The speed, $v=\sqrt{\summ v_i^2}$, like the fluctuating temperature, (\ref{eq:t}), is 
distributed as $\chi^2$, 
\begin{equation}
f_m(v)=\frac{2}{\left(2t_0\right)^{\mtwo}\Gamma(\mtwo)}
v^{m-1}e^{-v^2/2t_0}, \label{eq:Maxwell-gen}
\end{equation}
and their ratio 
\[
\frac{v^2}{mt}=\frac{\onem\summ v_i^2}{\onen\sumn v_i^2}, \]
is beta-distributed \cite{Cramer}. This is to say that since $v^2$ and $t$ are independent by hypothesis,
their ratio    will be distributed by the integrals of the product of their probability 
density functions (pdfs) over the domain defined by the inequalities $t>0$ and 
$0<v^2<mt x$. Introducing a new variable  $u=v^2/mt$, the probability
that $v^2/mt\le x$ is
\begin{eqnarray*}
\lefteqn{F_{m,n}(x)} \\
& = & a_{m,n}\int_{0}^x u^{\mtwo-1}\,du\int_{0}^{\infty}\,
t^{\mtwo+\ntwo-1}e^{-(n+mu)t/2t_0}\,dt, 
\end{eqnarray*}
where
\[a_{m,n}=\frac{m^{\mtwo}n^{\ntwo}}{\left(2t_0\right)^{\mtwo+\ntwo}
\Gamma(\mtwo)\Gamma(\ntwo)}.
\]
The integral over $t$ is a gamma integral so that
\[F_{m,n}(x)=\frac{m^{\mtwo}n^{\ntwo}}{B(\mtwo,\ntwo)}\int_{0}^x\,
\frac{u^{\mtwo-1}}{(n+mu)^{\mtwo+\ntwo}}du,\]
where $B(\cdot,\cdot)$ is the beta function. 
Differentiation then yields the pdf
\begin{equation}
f_{m,n}(x)=\frac{m^{\mtwo} n^{\ntwo}}{B(\mtwo,\ntwo)}
\frac{x^{\mtwo-1}}{(n+mx)^{\mtwo+\ntwo}},\;\;\;\;\; 0\le x<\infty
\label{eq:Fisher}
\end{equation}
which is known as the $F$-distribution in honor of its discover, R. A. Fisher.\par
It is quite remarkable that the pdf (\ref{eq:Fisher}) is independent
of the true temperature of the heat bath, $t_0$. The true temperature of the 
reservoir is irrelevant to the probability that $v^2/mt\le x$. 
There is a considerable chance that   values of the speed $v$ 
greater than $\sqrt{mt_0}$ are equally as likely  to be attributed to 
 values of the fluctuating temperature $t$ lower than $t_0$.
\par 
In the limit as $n\rightarrow\infty$, the $F$-distribution (\ref{eq:Fisher}) 
transforms into the $\chi^2$-distribution
\begin{equation}
f_{m,\infty}(x)=\frac{(\half m)^{\mtwo}}{\Gamma(\mtwo)}x^{\mtwo-1}e^{-\half mx}. 
\label{eq:chi}
\end{equation}
In particular, for $m=3$ and $x=\third(v^2/t_0)$, where $v=\sqrt{\sum_{i=1}^3v_i^2}$, 
 (\ref{eq:chi}) becomes the Maxwell speed distribution
\begin{equation}
f_{3,\infty}(v)=\sqrt{\frac{2}{\pi t_0^3}}v^2e^{-v^2/2t_0}
\label{eq:Maxwell}
\end{equation}
in $3D$ for a particle of unit mass. This shows that for a finite size, $n$,
the  temperature fluctuates, and, only asymptotically
tends to the temperature of the heat reservoir, $t_0=\beta^{-1}_0$, where $\beta_0$
 is the precision of the distribution \cite{Lindley}.\par
Tsallis and co-workers \cite{Tsallis} have attempted to generalize the Maxwellian, 
(\ref{eq:Maxwell})
in order to justify their variational procedure in maximizing the nonadditive
entropy
\begin{equation}
S_q(f(x))= \frac{1-\int\,dx f^q(x)}{q-1},\label{eq:Tsallis}
\end{equation}
with respect to the so-called escort probability averages \cite{BS}, \emph{e.g.\/},
\begin{equation}\int\,dx x^2f^q(x)=\left<x^2\right>_q.\label{eq:escort}
\end{equation}
In the case where the characteristic exponent $q>1$, they get the
 Student $t$-distribution \cite{Tsallis}
\begin{equation}
f(x)=\frac{\sqrt{\beta(q-1)}}{B(\half,\mbox{\small{$
\frac{1}{q-1}$}})}\frac{1}
{[1+\beta(q-1)x^2]^{1/(q-1)}}, \label{eq:t-Student}
\end{equation}
where $\beta$ is the Lagrange multiplier for the constraint (\ref{eq:escort}). 
In the case
where $q<1$ they obtain the Student $r$-distribution
\begin{equation}
f(x)=\frac{\sqrt{\beta(1-q)}}{B(\half,\mbox{\small{$\frac{1}{1-q}$}}+1)}
\left(1-\beta(1-q)x^2\right)^{1/(1-q)}. \label{eq:r}
\end{equation}
\par
It will be readily appreciated  that (\ref{eq:t-Student}) and (\ref{eq:r}) are
not pdfs for $x$ but only the product $\beta x^2$, or, equivalently, $x^2/t$  is the 
ratio of two $\chi^2$-variates. In other words, $\beta$ is not a
mere Lagrange multiplier for the constraint (\ref{eq:escort}), but, rather, is a random
variable itself. Since $\beta^{-1}=\onen\sumn v_i^2$, it will only converge in probability
to the absolute temperature in the limit as $n\rightarrow\infty$. And, in this limit,
both (\ref{eq:t-Student}) and (\ref{eq:r}) transform into $\chi^2$-distributions of
Maxwell-Boltzmann-Gibbs (MBG) statistics. Moreover,
we will now show that  the physical mechanisms leading to 
(\ref{eq:t-Student}) and (\ref{eq:r}) are  different, making it difficult to
believe that they can be derived from the same variational formalism.\par
The Student $t$-distribution (\ref{eq:t-Student}) is a particular case of the 
$F$-distribution,
\begin{equation}
f_{m,n}(z)=\frac{2m^{\mtwo}n^{\ntwo}}{B(\mtwo,\ntwo)}
\frac{z^{m-1}}{\left(mz^2+n\right)^{\mtwo+\ntwo}}, 
\label{eq:Fisher-bis}
\end{equation}
for $m=1$, where
\begin{equation}
z^2=\frac{\onem\summ v_i^2}{\onen\sumn v_i^2}.\label{eq:z}
\end{equation}
If $\beta^{-1}$ is to be identified as the  temperature \cite{Beck,Silva}, 
or the variance of the normal distribution $t_0$, then the limit $n\rightarrow\infty$ 
must be taken in 
(\ref{eq:Fisher-bis}), which gives the $1D$ Maxwell distribution. In other words,
$\beta^{-1}$ cannot be indentified with a fixed, constant temperature unless the 
number of degrees of freedom of the reservoir is allowed to increase without limit. But 
then the $F$-distribution reduces to the $\chi^2$-distribution with $m$ degrees of
freedom (\ref{eq:Maxwell-gen}), where we have set $z=v/\sqrt{mt_0}$.\par 
If we define the fractional energy as 
\[x=\frac{\summ v_i^2}{\summ v_i^2+\sumn v_i^2},\]
then $z^2=(n/m)x/(1-x)$, and the fractional energy, $x$, satisfies the beta pdf
\begin{equation}
f_{m,n}(x)=\frac{x^{\mtwo-1}(1-x)^{\ntwo-1}}{B(\mtwo,\ntwo)}.\label{eq:beta}
\end{equation}
Introducing the fluctuating temperature according to (\ref{eq:t}) and taking the limit
as $n\rightarrow\infty$ in (\ref{eq:beta}) gives the $\chi^2$-distribution 
(\ref{eq:Maxwell-gen}). \par
If we set $x$ equal to the fractional energy $E_1/E$, where $E_1$ is the energy of subsystem $1$ 
and $E_2=E-E_1$ is the energy of subsytem $2$, with $E$ as the total energy, then 
(\ref{eq:beta}) becomes the composition law for the structure function \cite{Khinchin}
\[\Omega(E)=\int_0^E\,\Omega_1(E_1)\Omega_2(E-E_1)\,dE_1=\frac{E^{\mtwo+\ntwo-1}}
{\Gamma(\mtwo+\ntwo)}.\]
The average energy of subsystem $1$ is
\begin{eqnarray*}
\lefteqn{\left<E_1\right>=}\\
& &\frac{1}{\Omega(E)}\int_0^E\,E_1\Omega_1(E_1)\Omega_2(E-E_1)\,dE_1
=\frac{m}{m+n}E.\end{eqnarray*}
The ratio of the expected energy  of subsystem $1$ to the total energy is just 
the fraction of degrees of freedom of subsystem $1$.
\par
However, no temperature can be defined in this microcanonical setting. 
Placing the system  in thermal contact with a heat resevoir of fixed temperature 
$\beta_0^{-1}$ is formally described by the Laplace transform 
\[\beta_0^{-\mtwo}=\int_0^\infty\,\frac{E_1^{\mtwo-1}}{\Gamma(\mtwo)}
e^{-\beta_0E_1}\,dE_1
\]
from which the normalized pdf follows
\begin{equation}
f_m(E_1|\beta_0)=\beta_0^{\mtwo}
\frac{E_1^{\mtwo-1}}{\Gamma(\mtwo)}e^{-\beta_0E_1}.\label{eq:exp}
\end{equation}
This gamma pdf for the energy of subsystem $1$ becomes the $\chi^2$-distribution
(\ref{eq:Maxwell-gen}) under the subsitution $E_1=\half v^2$. The canonical ensemble
has built into it the asymptotic limit of an infinite heat reservoir, consisting
of subsystem $2$, so that a fluctuating temperature cannot   be defined in a 
\emph{frequency\/} sense. However,
the temperature of the heat bath is usually unknown, and it must be inferred by making
measurements on the energy of the small subsystem $1$. Therefore, the estimator of
the inverse temperature, $\beta$, becomes a function of the sample values of the energy
and because the latter fluctuate, the former will also fluctuate. The pdf of
$\beta$ will not be in the frequency sense, but, rather, in the sense of degree of
belief that certain values of $\beta$ are more likely than others \cite{Lavenda}.\par
For instance, Bayes's theorem, giving the posterior pdf of $\beta$ as
\begin{equation}\tilde{f}(\beta|E_1)= \frac{f(E_1|\beta)\tilde{f}(\beta)}
{\int_0^\infty\,f(E_1|\beta)\tilde{f}(\beta)\,d\beta},\label{eq:Bayes}
\end{equation}
where $\tilde{f}$ represents a pdf in the sense of degree of belief
\cite[p. 1]{Lindley}, would give a $\chi^2$-distribution
for $\beta$ instead of $t$, as in (\ref{eq:t-dis}), if Jeffreys's rule for the
prior pdf  $\tilde{f}(\beta)$ is employed \cite[eqn (4.97)]{Lavenda}:
\begin{equation}
\tilde{f}(\beta|E_1)=E_1^{\mtwo}\frac{\beta^{\mtwo-1}}{\Gamma(\mtwo)}
e^{-\beta E_1}.\label{eq:exp-bis}
\end{equation}
Whereas (\ref{eq:exp}) is a gamma pdf for $E_1$, (\ref{eq:exp-bis}) is a gamma pdf
for $\beta$. This is in  contradistinction to the $\chi^2$-distribution
for the temperature $t$, (\ref{eq:t-dis}). In fact, (\ref{eq:exp-bis}) implies
that the temperature be distributed according to an inverse gamma pdf.\par Bayes's theorem (\ref{eq:Bayes})
tells us how the pdf of the beliefs about $\beta$ is modified from our previous 
knowledge, contained in $\tilde{f}(\beta)$, as a result of the measurement, $E_1$. For
the exponential family of pdfs, (\ref{eq:exp}), even a single measurement, $E_1$, is 
a \emph{sufficient\/} statistic. When little is known other than the parameter's value 
can have any value from $0$ to $\infty$, Jeffreys's rule is to take its logarithm as uniformly distributed. 
\par
In contrast to the Bayes's interpretation of $\beta$ in terms of degree of belief, the 
fluctuating temperature (\ref{eq:t}) is to be interpreted in the
frequency sense, since it is distributed as $\chi^2$ in (\ref{eq:t-dis}). The average 
of the mean of $v_i^2$ found from (\ref{eq:Maxwell-gen})
\begin{equation}
\onem\left<\summ v_i^2\right>=t_0=\lim_{n\rightarrow\infty}\onen\sumn v_i^2
 \label{eq:equi}
\end{equation}
gives the equipartition result. The second equality in (\ref{eq:equi}) is the definition
of the absolute temperature, and it asserts that the arithmetic mean of the squares of the 
velocities of all
the particles in the reservoir, in the limit as their number tends to increase
without limit, is equal to the arithmetic mean of the sum of the expectation 
of the squares of the velocities of the finite number of particles comprising the
system. The latter can also be equal to the number of degrees of freedom of a single
particle, $m=3$.\par
The estimator (\ref{eq:t}) is said to be a \emph{consistent\/} insofar as $n$
increases indefinitely it converges in probability to the value (\ref{eq:equi})
\cite[p. 351]{Cramer}. This is guaranteed by
the strong law of large numbers: $t$ is the mean of random variables, $v_i^2$, which are
independent and normally distributed \cite[p. 30]{Lindley}. The $\chi^2$-distribution
(\ref{eq:t-dis}) shows that $t$ is an \emph{unbiased\/} estimator in that 
$\left<t\right>=t_0$, whatever the value of $n$. The variance of (\ref{eq:t-dis}), 
$2t_0^2/n$, shows that $t$ is an \emph{efficient\/} estimator.
\par
The Student $r$-distribution (\ref{eq:r}) reflects a different physical process for 
defining the \lq temperature\rq\ of a particle. All the particles of the system now 
comprise the reservoir, including the test particle under examination. Consequently,
the numerator and denominator of
\[
r=\frac{v_1}{\sqrt{t}}=\frac{v_1}{\sqrt{\onen\sumn v_i^2}} \]
are not independent so that the previous method used to derive the $F$-distribution 
(\ref{eq:Fisher-bis}),
in general, and the Student $t$-distribution (\ref{eq:t-Student}), in particular, 
will not work.\par However, by defining the variable \cite[p. 240]{Cramer}
\[y=\sqrt{\frac{n-1}{n}}\frac{r}{\sqrt{1-\onen r^2}}=\frac
{v_1}{\onenone\sumntwo v_i^2},\]
the numerator and denominator become independent of each other. Then since $y$ is 
distributed according to the Student $t$-distribution with $n-1$ instead of $n$, $r$
will be distributed according to
\begin{equation}f_n(r)=\frac{1}{\sqrt{n}}\frac{1}{B(\half,\mbox{\small{$\frac{n-1}{2}$}})}
\left(1-\frac{r^2}{n}\right)^{\half(n-3)},\label{eq:r-bis}
\end{equation}
which is the Student $r$-distribution (\ref{eq:r}), where $|r|\le\sqrt{n}$. Although
Tsallis and co-workers derive (\ref{eq:r-bis}) from their variational formalism
for $q<1$, it involves a different physical process in comparison to the $t$-distribution, 
(\ref{eq:t-Student}), which they get for $q>1$.\par Since
$v_1$ can be both negative and positive, the $r$-distribution will be $U$-shaped, 
like the arcsine distribution, for $n=2$, having a minimum at the mean 
$\left<r\right>=0$. For $n=3$, it
becomes the rectangular distribution since $-\sqrt{n}\le r\le\sqrt{n}$. While, for
$n>3$, the $r$-distribution becomes unimodal with mean zero and variance $\left<
r^2\right>=1$, for all values of $n$. Only in the asymptotic limit as $n\rightarrow
\infty$ does the $r$-distribution transform into the single particle Maxwellian, where
\begin{equation}
\left<v_1^2\right>=t_0=\lim_{n\rightarrow\infty}\onen\sumn v_i^2. \label{eq:equi-bis}
\end{equation}
In comparison with (\ref{eq:equi}), (\ref{eq:equi-bis}) reflects only the single 
particle, or single degree of freedom, nature of the equipartition law.\par
Alternatively, if we allow $m$ to increase without limit, while keeping $n$ finite and fixed, the
 $F$-distribution (\ref{eq:Fisher-bis}) transforms into the inverted gamma pdf
\begin{equation}
f_{\infty,n}(z)=\frac{2}{\Gamma(\ntwo)}\left(\ntwo\right)^{\ntwo}\frac{1}
{z^{n+1}}e^{-n/2z^2}, \label{eq:ig}
\end{equation}
since 
\[\lim_{m\rightarrow\infty}\frac{1}{m^{\ntwo}}\frac{\Gamma(\mtwo+\ntwo)}{\Gamma(\mtwo)}
=\frac{1}{2^{\ntwo}}.\]
However, on consulting (\ref{eq:z}) in the limit as $m\rightarrow\infty$, we find
$z=\sqrt{nt_0}/v$. Upon introducing this transformation into the inverted gamma pdf
(\ref{eq:ig}) we get the $\chi^2$-distribution (\ref{eq:Maxwell-gen}) with $n$ in place
of $m$. Hence, even though the Fisher pdf (\ref{eq:Fisher-bis}) is not symmetric in the
degrees of freedom, $m$ and $n$, the asymptotic limits are the same, \emph{i.e.\/}, 
$f_{\ell,\infty}(v)=f_{\infty,\ell}(v)$, for systems of $\ell$ degrees of freedom.
\par
On the basis of the foregoing analysis we conclude that temperature, which is the 
true value or the variance of the distribution (\ref{eq:Maxwell-gen}), can only be introduced into MBG statistics, since the random variables,
$v_i^2$ are independent and normally distributed, and their mean tends to the
absolute temperature, $t_0$, by the strong law of large numbers. The 
fluctuating temperature (\ref{eq:t}) is distributed according to the 
$\chi^2$-distribution (\ref{eq:t-dis}). This, however,  is in blatant contradistinction with the identification of $\beta$  as a 
Lagrange multiplier in the derivation of the Student $t$- and $r$- distributions from 
the variational formalism that seeks to maximize the nonadditive entropy (\ref{eq:Tsallis})
with respect to the escort probability averages, (\ref{eq:escort}).
\par


\begin{thebibliography}{99}
\bibitem{Cramer}H. Cram\'er, \emph{Mathematical Methods of Statistics\/} (Princeton
U. P., Princeton NJ, 1946), pp. 241--243.
\bibitem{Lindley}D. V. Lindley, \emph{Introduction to Probability and Statistics\/},
part 2,
(Cambridge U.P., Cambridge, 1970), p. 8.
\bibitem{Tsallis}A. M. C. de Souza and C. Tsallis, \emph{Physica A\/} \textbf{236}
(1997) 52--57.
\bibitem{BS}C. Beck and F. Schl\"ogl, \emph{Thermodynamics of Chaotic Systems\/}
(Cambridge U. P., Cambridge, 1993), p. 53.
\bibitem{Beck}C. Beck, \lq\lq Dynamical foundations of nonextensive statistical
mechanics\rq\rq\ (cond-mat/0105374).
\bibitem{Silva}R. Silva, A. R. Plastino and J. A. S. Lima, \lq\lq A Maxwellian
path to the $q$-nonextensive velocity distribution function\rq\rq\ (cond-mat/0201503).
\bibitem{Khinchin}A. I. Khinchin, \emph{Mathematical Foundations of Statistical
Mechanics\/} (Dover, New York, 1949), p. 75.
\bibitem{Lavenda}
B. H. Lavenda, \emph{Statistical Physics: A Probabilistic
Approach\/} (Wiley-Interscience, New York, 1991), \S$\;$ 4.3 and \S$\;$ 4.11.
\end{thebibliography}
\end{document}